\documentclass{ws-procs9x6}

\begin{document}


\title{A few words on Entropy, Thermodynamics, and Horizons}

\author{Donald Marolf}
\address{Physics Department \\
University of California, Santa Barbara USA 93106 \\
marolf@physics.ucsb.edu}


\maketitle

\abstracts{We review recent progress in understanding certain
aspects of the thermodynamics of black holes and other horizons.
Our discussion centers on various ``entropy bounds'' which have
been proposed in the literature and on the current understanding
of how such bounds are {\it not} required for the semi-classical
consistency of black hole thermodynamics.  Instead, consistency
under certain extreme circumstances is provided by two effects.
The first is simply the exponential enhancement of the rate at which a macrostate with large entropy is emitted
in any thermal process.  The second is a
new sense in which the entropy of an ``object'' depends on the
observer making the measurement, so that observers crossing the
horizon measure a different entropy flux across the horizon than
do observers remaining outside.  In addition to the review, some recent criticisms are addressed.  In particular, additional arguments and detailed numerical calculations showing the observer dependence of entropy are presented in a simple model.  This observer-dependence may have further interesting implications for the
thermodynamics of black holes.  
For the Proceedings of the GR17 conference, Dublin, Ireland, July 2004.
}


\section{Introduction}

Most researchers agree that black holes, Hawking radiation, and
black hole entropy are some of our strongest clues in the quest to
discover and understand ever deeper layers of fundamental physics.
However, somewhat less agreement has been reached as to the nature
of the fundamental physics to which they will lead. Below, we
address only a narrow part of this discussion, reviewing recent
work \cite{MS1,MS2,MMR} which exposes loopholes in certain
arguments for so-called ``entropy bounds''--
conjectures\cite{Bek73,tH,LS,FS,Bousso,BV} such as Bekenstein's
proposed bound\cite{Bek73},
\begin{equation} \label{Bek}
  S < \alpha RE/\hbar c,
\end{equation}
and the so-called holographic bound\cite{tH,LS},
\begin{equation}
\label{holog}
  S < A c^3 / 4 \hbar G,
\end{equation}
that would require the entropy of any thermodynamic system to
respect a bound set by the size and, perhaps, the energy of the
system\footnote{Here we have displayed the fundamental constants
explicitly, but below we use geometric units with $k_B
=\hbar=c=G=1$. The original version\cite{Bek73} of Bekenstein's
bound has $\alpha=2\pi$, while some subsequent discussions (e.g.
Ref. \cite{Erice}) weaken the bound somewhat, enlarging $\alpha$
by a factor of order ten.}. We also review an unexpected piece of
new physics that forms the basis of one such loophole: a new
dependence\cite{MMR} of entropy on the observer, and in particular
a dependence on the observer of the flux of entropy across a null
surface.  This aspect of the review is supplemented below with a detailed new analysis of a simple example, using both analytic and numerical methods.

To place this review in its proper context, recall that the above
conjectured entropy bounds were originally motivated by claims
that they are required for the consistency of black hole entropy
with the second law of thermodynamics.  These claims have been
controversial\cite{UW,UW2} since they were first introduced. However, last year
a new class of very general loopholes was pointed out in Refs.
\cite{MS2,MMR}.  We review and comment on these loopholes below.

While these works still have not quieted all controversy\cite{Rumanov,newBek}, they have gone some distance toward doing so.  In particular, we will see that the criticisms of Ref. \cite{Rumanov} (which were stated in the context of a certain simple model) can be refuted in complete detail.   This is done twice in section \ref{calcs}, once using analytic arguments and once numerically.  The numerical results are an excellent match to the analytic predictions.   
On the other hand, as discussed in section \ref{genC}. the criticisms of Ref. \cite{newBek} apply only to a special case.  Furthermore, even in that context, Ref. \cite{newBek}  makes an assumption about the nature of Hawking radiation which appears to conflict with well-established results\cite{Hawking,KW,FH}. Thus, the original loopholes\cite{MS2,MMR} stand firm.

While the discovery of loopholes does not prove the conjectured
entropy bounds of  Refs. \cite{Bek73,tH,LS,FS,Bousso} to be false\footnote{In particular, there are certain observational bounds (see, e.g. Ref. \cite{MS2}) which limit the existence of highly entropic objects in our universe.  While the form of such observational bounds does not obviously match that of (\ref{Bek}) or (\ref{holog}) it would be interesting to explore such observational bounds in more detail.  Indeed, Bekenstein stresses that his interest in the proposal (\ref{Bek}) is in the context of our particular universe.  In contrast, we are more interested here possible deep ties between the bounds and quantum gravity; i.e., in the question of whether (\ref{Bek}) and (\ref{holog}) could in principle be violated in some consistent theory of quantum gravity, which may or may not describe our particular universe.   We will therefore discuss (\ref{Bek}) and (\ref{holog}) in this strong sense below.} it does remove the original motivation for such bounds. Proponents
may still appeal to the AdS/CFT correspondence\cite{Juan} and in
particular to the Susskind-Witten calculation\cite{SW} for support, as well as
the difficulty (see e.g. Refs. \cite{FMW,RaphRev,BFM}) of
constructing counter-examples to the Covariant Entropy Bound
conjecture\cite{Bousso} within the particular regime in which it
has been formulated.  However, the lessons of these observations
may in the end turn out to be more subtle, and it is clear that no
compelling evidence for the bounds is available at this time.

The plan of this review is as follows.  We first outline the
loopholes in the arguments of Refs.
\cite{Bek73,tH,LS,Erice,Bek2000}.   Our treatment (section \ref{loop}) roughly
follows that of Ref. \cite{MS2}, and more details can be found in
that reference. Section \ref{obs} then follows Ref. \cite{MMR} and
moves on to consider a related thought experiment, which one might
at first think would require similar restrictions on matter for
the consistency of the second law. However, a careful examination
of the physics in fact leads to a different conclusion, in which
the issue is resolved by the observer-dependence of entropy foreshadowed above. 

\section{Thermodynamics can hold its own}

\label{loop}

Below, we turn our attention first to bounds of the form
(\ref{Bek}), and then to bounds of the form (\ref{holog}).  In
each case, the essential ingredient is the well
established\cite{Hawking,UW2,Ray,KW,BFZ,TJ1,TJ2,TJ3} point that the radiation
surrounding a black hole of temperature $T_{BH}$ is thermal in the
sense that, {\it in equilibrium}, it is described by an ensemble
of the form $e^{-\beta H}$, where $\beta=1/T_{BH}$.   As a result,
the probability to find a particular {\it macrostate} (such as a
bound-violating box) in a thermal ensemble is not $e^{-\beta E}$
but $e^{-\beta F}$, with $F=E-T_{BH}S$ the free energy of the
macrostate at the black hole temperature $T_{BH}$ and where the
phase space factor of $e^S$ arises from associating  $e^S$
microstates with a single macrostate.

\subsection{General Comments}
\label{genC}

Of interest here will be the somewhat different setting in which a black
hole radiates into empty space.  It is more difficult in this context to arrive at detailed results for interacting fields, though for free fields one can readily show\cite{Hawking,FH} that the radiation produced by the black hole is just the outgoing component of the above thermal ensemble.  This has been established rigorously\cite{FH} even for the complicated case where the black hole forms dynamically from the collapse of some object.  

Consider now the case of interacting fields.  Suppose in particular that  the thermal ensemble can be described as a collection of weakly interacting particles or quasi-particles.   We would again like to claim that the outward flux of both particles and quasi-particles from a black hole radiating into free space is well-modeled by the outgoing component of a thermal ensemble.  Let us therefore consider placing our black hole inside spherical box whose walls are held at a fixed temperature.  In this context, the above results tell us that the black hole will reach eventually reach thermal equilibrium.  Since the particles and quasi-particles interact weakly, to a good approximation we may regard this ensemble as a combination of its ingoing and outgoing parts.  Since the ingoing parts incident on the black hole are completely absorbed,  equilibrium can only be maintained if the black hole also radiates a flux of both particles and quasi-particles at a rate given by the outgoing part of the thermal ensemble.  As discussed in the conclusions of Ref. \cite{FH}, even in the presence of strong interactions one expect that the outgoing Hawking radiation is determined by properties of the thermal ensemble, though the details in that case will be quite complicated.  Thus, any proposal (such as that of Ref. \cite{newBek}) that Hawking radiation of quasi-particles be treated differently that of ``fundamental'' particles, and which results in interesting cases in an outward flux of quasi-particles significantly lower than which arises in equilibrium is in conflict with standard results\cite{Hawking,UW2,Ray,KW,BFZ,TJ1,TJ2,TJ3}.  

Because just such a reduced quasi-particle model (in the context of weak interactions) formed the basis\footnote{We also note below that quasi-particles are not essential for the loopholes we discuss.  Thus, it is not strictly necessary to refute Bekenstein's recent comments\cite{newBek}.  However, we do so here in order to show that the loopholes are quite general.} of a recent criticism\cite{newBek} by Bekenstein of the work\cite{MS2} to be described below, it is worthwhile to examine this point once more in great detail.  The particular proposal of Ref. \cite{newBek} was that quasi-particles cannot be {\it directly} Hawking radiated by a black hole, but instead arise in thermal equilibrium only via additional processes in which they are created by the interaction of other (so-called ``fundamental'') particles.  The case of interest\cite{newBek} is one where kinetics would dictate that such a process happens only extremely slowly, on some timescale which we will call $t_{quasi}$.  But now consider the case of a black hole in Anti-de Sitter (AdS) space.  Here we know\cite{KW,BFZ,TJ1,TJ2,TJ3} that there is a state (effectively, the Hartle-Hawking vacuum) in which the black hole is at thermal equilibrium.  Yet if the interactions are weak, any object with zero angular momentum will fall into the black hole 
on a timescale $t_{AdS}$ set by the size of the AdS space.  Thus, under the proposal of Ref. \cite{newBek}, one may choose $t_{quasi} \gg t_{AdS}$ so that the quasi-particles drain out of the ensemble and into the black hole far faster than they are created.  Thus there could be no thermal equilibrium, contradicting in particular the theorems of Ref. \cite{KW}.  One can only conclude that the reduced-quasi-particle proposal\cite{newBek} is untenable, and that quasi-particles are directly Hawking radiated by the black hole at a rate set by the outgoing flux of such objects in the thermal ensemble determined by the black hole's Hawking temperature.

\subsection{The Bekenstein bound}
\label{Bb}

Let us now recall the setting for the argument of Ref.
\cite{Erice} in favor of (\ref{Bek}). One considers an object of
size $R$, energy $E$, and entropy $S$ which falls into a
Schwarzschild black hole of size $R_{BH}$ from a
distance $d{\gg}R_{BH}$. Following Ref. \cite{Erice}, we parametrize the problem by $\zeta = \frac{R_{BH}}{2R}$.  The parameter $\zeta$ is taken to be
large enough that the object readily falls into the black hole
without being torn apart. In other words, we engineer the
situation so that the black hole is, at least classically, a
perfect absorber of such objects. It is also assumed that the
Hawking radiation emitted during the infall of our object is
dominated by the familiar massless fields. Assuming that the
number of such fields is not overly large, the effects of Hawking
radiation are negligible during the infall.  Under these assumptions,  Ref. \cite{Erice} shows that
the second law is satisfied only if
\begin{equation}
                 S < 8\pi \nu \zeta R E,
\end{equation}
where $\nu$ is a numerical factor in the range $\nu=1.35-1.64.$
Here the energy $E$ has been assumed to be much smaller than the
mass $M_{BH}$ of the black hole and, up to the factor $\nu$, the
above bound is obtained by considering the increase in entropy of 
the black hole, $dS_{BH}=dE_{BH}/T_{BH}$. Notice here that the
black hole was in particular assumed not to readily emit copies of our object as
part of its Hawking radiation.

However, let us now suppose that a ``highly entropic object'' can
exist with $S > 8\pi \nu \zeta R E$.  Then, according
to Ref. \cite{Erice}, this will lead to a contradiction with the second law
in a process involving the particular black hole on which we have
chosen to focus. We wish to consider the possibility that, because
this entropy is very large, such objects {\it will} be readily emitted
by the black hole so that the Hawking radiation will {\it not} be
dominated by massless fields. In particular, if they are emitted
sufficiently rapidly, the black hole will emit such objects faster
than we have proposed to drop them in and, as a result, it is
clear that the second law will be satisfied.  It is natural to consider such highly entropic objects as being built of many parts and thus, in some sense, ``composite'' so that they may best be described as long-lived quasi-particle excitations of some fundamental theory.  However, for clarity we point out that such complications are unnecessary, and the reader is encouraged to keep in mind the concrete example of a system with $e^S$ species of fundamental massive scalar field, all having of mass $E$.  In this example, the entropy arises if one takes the fields to have sufficiently similar interactions such that it is reasonable to coarse-grain over the species label.

To study the production rate, we follow the path set out in section \ref{genC} above and compute the free energy of our object at the black hole temperature $T_{BH}=(4\pi
R_{BH})^{-1}=(8\pi\zeta R)^{-1}$.  \ We have
\begin{equation}
            F = E - \frac{S}{4 \pi R_{BH}} < E - \nu E < 0.
\label{Feqn}
\end{equation}
Thus, we expect an exponentially {\it large} rate of Hawking
emission of these objects  from this black hole.

In fact, in a naive model the resulting thermal ensemble contains
a sum over states with arbitrary numbers $n$ of such highly
entropic objects, weighted by factors of $e^{-nF} > 1$.  Such an
ensemble clearly diverges and is physically inadmissible.  Thus, some effect must intervene to cut off the
divergence and to stabilize the system.  If no other effect
intervenes first, this cut-off can be supplied by the quantum
statistics of the highly entropic objects.  To see this, we note
that the above naive model treats the objects as {\it
distinguishable}.  In contrast, the partition function for $N$
species of either boson or fermion converges at any temperature no
matter how large $N$ may be.  Thus, if our ``object" were a
collection of $N = e^S$ free Bose or Fermi fields, the associated
partition function would converge.  The divergence of the naive
model simply indicates that the true equilibrium ensemble contains
so many boxes that they cannot be treated as {\it distinguishable}
and {\it non-interacting} objects! So long as there is no strong
barrier that prevents the Hawking radiation from escaping, the
outgoing radiation will also be correspondingly dense, at least in
the region near the black hole.

On the other hand, suppose that the boxes interact in such a way that the thermal
atmosphere of the black hole largely blocks the passage of outward-moving boxes.
Let us assume that it also blocks
the passage of the CPT conjugate objects, since these will carry
equal entropy.  This might happen because the thermal
atmosphere already contains many densely packed copies of our
object, or alternatively because our objects are blocked by other components of the atmosphere.
However, (due to CPT invariance), such an atmosphere
will also provide a barrier hindering our ability to drop in a new object from far away.
If we attempt to drop in such an object, then with correspondingly large probability it must bounce off the thermal atmosphere or be
otherwise prevented from entering the black hole!  Again, one expects that the black hole radiates at least one object for every highly entropic object which is successfully sent inward through the
atmosphere from outside.

The argument above was stated in terms of the particular scenario
suggested in Ref. \cite{Erice}.  However, one might expect that it
can be formulated much more generally. This is indeed possible, as
may be seen by considering any process in which a given object
(``box" above) with entropy $S$ is hidden inside the black hole, with the black hole receiving its energy $E$ and its entropy $S$ otherwise disappearing from the universe.  Note that this includes both processes of the
original form\cite{Bek73} where the object is slowly lowered into
the black hole,  as well as the more recent version\cite{Erice} discussed above
where the object falls freely. As before, we suppose that this
represents a small change, with $E$ being small in comparison to
the total energy of the black hole. From the ordinary second law
we have
\begin{equation}
   \Delta S_{total} \ge \Delta S_{BH} - S,
\end{equation}
but the {\it first} law of black hole thermodynamics tells us that
this is just
\begin{equation}
\Delta S_{total} \ge \frac{E}{T_{BH}} - S = \frac{F}{T_{BH}},
\end{equation}
where $F$ is the free energy of the object at the Hawking
temperature $T_{BH}$. In particular, since $T_{BH}>0$, the sign of
$\Delta S_{total}$ must match that of $F$. Thus absorption of an
object by a black hole violates the second law only if $F<0$, in
which case any of the above mechanisms may either prevent the process
from occurring or sufficiently alter the final state so that the second law is satisfied.

\subsection{The holographic bound}
\label{hologObs}

We now turn to the ``holographic bound'' (\ref{holog}) and
consider the analogous arguments of Refs. \cite{tH,LS}. These
works suggest that one compare a (spherical, uncharged) object
with $S \ge A/4$ with a Schwarzschild black hole of equal area $A
= 4 \pi R_{BH}^2$. Since the highly entropic object is not itself
a black hole, its energy $E$ must be less than the mass $M_{BH}$
of the black hole. One is then asked to drop the bound-violating
object into a black hole of mass $M_{BH}-E$ or otherwise transform
it into a black hole of mass $M_{BH}$. For arguments which drop
the object into a pre-existing black hole, one
typically\footnote{See, e.g., the weakly gravitating case
described in Ref. \cite{Bek2000}.} (though not always) assumes $E
\ll M_{BH}$.

Let us suppose for the moment that $E \ll M_{BH}$.  Then we may
estimate the emission rate of such `highly entropic objects' from
a black hole of mass $M_{BH}$ by considering the free energy
$F=E-TS$ of our object at the Hawking temperature $T_{BH} = (4 \pi
R_{BH})^{-1}$. We have
\begin{equation}
\label{hF} F = E - T_{BH} S  < E - \frac{A/4}{4 \pi R_{BH}} = E  -
M_{BH}/2 < 0.
\end{equation}
Thus, we again see that our object is likely to be emitted readily
in Hawking radiation.  As a result, one may also expect
significant radiation during the formation of the black hole so
that no violation of the 2nd law need occur.

And what if $E$ and $M_{BH}$ are comparable?  Then back-reaction
will be significant during the emission of one of our objects from
the black hole and reliable information cannot be obtained by
considering an equilibrium thermal ensemble at fixed temperature.
 In fact, we are now outside the domain of validity of Hawking's calculation\cite{Hawking}.
However, Refs. \cite{kpw,mp} suggest how an emission rate may be
estimated in this regime.  These works find that the emission rate
for a microstate is proportional to $\exp(\Delta S_{BH})$, where
$\Delta S_{BH}$ is the change in black hole entropy when the
particle is emitted.  Note that this change is caused by the loss
of energy by the black hole and is typically negative.  Following
this conclusion allows us to proceed in much the same way as in
section \ref{Bb}: Forming a black hole from our object would
violate the second law only if $S>S_{BH}$, but since $\Delta
S_{BH}>-S_{BH}$ we have $\Delta{S}_{total}=\Delta{S}_{BH}+S>0$ for
the corresponding emission process.  Thus, Refs. \cite{kpw,mp}
predict an exponentially large emission rate for our object.   We
conclude that the combined process of collapse and emission would
actually result in a net {\it increase} in the total entropy.

\section{The Observer Dependence of Entropy}
\label{obs}

In section \ref{loop} above we described loopholes in certain
arguments\cite{Bek73,tH,LS,Bek2000,Erice} which appeared to suggest that
(\ref{Bek}) and (\ref{holog}) might be required if the 2nd law is
to be satisfied in general processes involving black holes.  In
particular, we argued that during the time it takes to insert an
object violating (\ref{Bek}) into a black hole of the appropriate
size, the black hole will in fact radiate many similar objects.
The result with then be a net production of entropy.

It turns out to be instructive to carry this analysis one step
further.  Following Ref. \cite{MMR}, let us therefore consider a
related thought experiment in which we first place our black hole
in a reflecting cavity (or, perhaps, in a small anti-de Sitter
space) and allow it to reach thermal equilibrium.  Note that in
equilibrium the rate at which highly entropic objects are absorbed
by the black hole exactly balances the emission rate.  But what
happens if we now alter the state of the system by adding yet one
more highly entropic object, say, violating (\ref{Bek}) and
travelling inward toward the black hole?

To simplify the problem, we take the limit
of a large black hole.  We may then describe the region near the
horizon by Rindler space.  Recall\cite{JP} that an object crossing
a Rindler horizon is associated with a corresponding increase in
the Bekenstein-Hawking entropy of this horizon by at least $\Delta
S \ge E_{acc}/T$ where $E_{acc}$ (the energy of the object as
measured by the accelerated observer) is the Killing energy
associated with the boost symmetry $\xi$ (i.e., the Rindler time
translation) and the associated temperature is given by
$\kappa/2\pi$ where $\kappa$ is the surface gravity of $\xi$.  As
usual, the normalization of $\xi$ cancels so that $E_{acc}/T$ is
independent of this choice.  Thus, in this setting the interesting
highly entropic objects are those with entropy $S > E_{acc}/T$.
Might these lead to a failure of the 2nd law for the associated
observers when such an object falls across the horizon?  Note that
because uniformly accelerated observers perceive the Minkowski
vacuum as a state of thermal equilibrium at temperature $T \neq
0$, this setting is analogous to the query from the previous
paragraph involving black holes.

We now proceed to analyze this situation following Ref. \cite{MMR}.
Let us first review the inertial description of this process, for which the state with no objects present is
the Minkowski vacuum  $|0_M\rangle$. The presence of an object
is then described as an excitation of this vacuum.  If there are $n$
possible microstates $|1_M;i \rangle$ ($i \in \{1,...,n\}$) for
one such object, then an object in an {\it undetermined}
microstate is described by the density matrix
\begin{equation}
\rho_M = {1 \over n} \sum_{i=1}^n |1_M;i \rangle \langle1_M;i|.
\end{equation}
The entropy assigned to the object by the inertial observer is
then as usual
\begin{equation}
S_M = - \mbox{Tr} \rho_M \ln \rho_M = \ln n.
\end{equation}

Of course, the inertial observer can still access the object (and
its entropy) after the object crosses the horizon so that there is
no possibility of a 2nd law violation from the interital point of
view.  Thus, the interesting question is what entropy a Rindler
observer assigns to this object. In particular, we wish to compute
the change in the amount of entropy accessible to the Rindler
observer when the object falls across the horizon.  This is just
the difference between the entropies of the appropriate two states
given by the Rindler descriptions of $\rho_M$ and $|0_M \rangle$.
An important point is that, since $|0_M \rangle$ is seen as a
thermal state, it already carries a non-zero entropy that must be
subtracted.   Similarly $E_{acc} = \delta E$ is the corresponding
difference in Killing energies.   We will work in the approximation
where gravitational back reaction is neglected and, in particular,
in which the horizon is unchanged by the passage of our object.

Following Ref. \cite{MMR}, let us consider then the thermal
density matrix $\rho_{R0}$ which results from tracing the
Minkowski vacuum, $|0_M\rangle \langle 0_M|$, over the
invisible\footnote{The terms "visible" and "invisible" Rindler
wedge will always be used in the context of what is in causal
contact with our chosen Rindler observer.  Of course, the entire
spacetime is visible to the inertial observer.} Rindler wedge:
\begin{equation}
\rho_{R0} = \mbox{Tr}_{invisible} |0_M \rangle \langle 0_M |.
\end{equation}
This describes all information that the Rindler observer can
access in the Minkowski vacuum state. We wish to compare
$\rho_{R0}$ with another density matrix $\rho_{R1}$ which provides
the Rindler description of the state $\rho_M$ above:
\begin{equation}
\rho_{R1} = \mbox{Tr}_{invisible} \rho_M = \mbox{Tr}_{invisible}
\frac{1}{n} \sum_{i=1}^n |1_M;i \rangle \langle1_M;i|.
\end{equation}

We would like to compute the difference in energy
\begin{equation}
\delta E= \mbox{Tr} [H(\rho_{R1}-\rho_{R0})],
\end{equation}
and in entropy
\begin{equation}
\delta S = - \mbox{Tr} [\rho_{R1} \ln \rho_{R1} - \rho_{R0} \ln
\rho_{R0}],
\end{equation}
where $H$ is the Hamiltonian of the system and, in both cases, the
sign has been chosen so that the change is positive when
$\rho_{R1}$ has the greater value of energy or entropy.

Now, the entropy of each state separately is well known to be
divergent\cite{entangle}.  However, Ref. \cite{MMR} points out
that the change $\delta S$ will still be well-defined.  In
particular, suppose the object has some moderately
well-defined energy $E_{inertial}$ as described by the inertial
observer and where the object is well localized.  Then the energy
measured by the Rindler observer will also be reasonably
well-defined and the difference $\rho_{R1} \ln \rho_{R1} -
\rho_{R0} \ln \rho_{R0}$ will have negligibly small diagonal
entries at high energy.  Thus, the above trace will exist. In
other words, we may compute $\delta S$ by first imposing a cutoff
$\Lambda$, computing the entropy ($S_0,S_1$) and energy
($E_0,E_1$) of the two states ($\rho_{R0},\rho_{R1}$) separately,
subtracting the results, and removing the cutoff. 

Now, the case which is most interesting for
the second law is where $\ln n \gg E_{acc}/T$ so that the presumed
violation of the second law is very large.  Note that in this
regime the expected number of similar objects
in the thermal ensemble $\rho_{R0}$ is already large; in particular, for $E_{acc} \gg T$ it is of order $n e^{-E_{acc}/T}$.    The
physical insight of Ref. \cite{MMR} is that, in such a case, the
difference between $\rho_{R1}$ and $\rho_{R0}$ is roughly that of
adding {\it one}\footnote{Actually, a number $\delta$ which depends on $E_{inertial}$ and $T$ but is, at least in simple cases, independent of $n$; see, e.g., eq. (\ref{enhance}) below and the surrounding discussion.} object to a collection of $n e^{-E_{acc}/T}$
similar objects.  Thus, one should be able to describe $\delta
\rho = \rho_{R1} - \rho_{R0}$ as a small perturbation.  One may
therefore approximate $\delta S$ through a first-order Taylor
expansion around $\rho_{R0}$:
\begin{equation}
\label{Sresult} \delta S \approx - \mbox{Tr}\left[ \delta \rho
{\delta (\rho \ln \rho)
    \over \delta \rho}\Big|_{\rho=\rho_{R0}} \right] = - \mbox{Tr} [
    \delta \rho (1 + \ln \rho_{R0})] = - \mbox{Tr} [\delta \rho \ln
    \rho_{R0}],
\end{equation}
where in the last step we use the fact that $\mbox{Tr} \rho_{R1} =
\mbox{Tr} \rho_{R0}$. Now recall that the initial density matrix
is thermal, $\rho_{R0} = e^{-H/T}/(\mbox{Tr} e^{-H/T})$.  Thus, we have
\begin{equation}
\delta S \approx - \mbox{Tr} [ \delta \rho (-H/T)] = {\mbox{Tr} [H
    (\rho_{R1} - \rho_{R0})] \over T} = {\delta E \over T},
\end{equation}
where we have again used $\mbox{Tr}(\delta \rho)=0$.  As noted in
Ref. \cite{MMR}, this key result is independent of any cut-off.

The above result can also be understood on the basis of classical
thermodynamic reasoning. The initial configuration $\rho_{R0}$
represents a thermal equilibrium. We wish to calculate the change
in entropy during a process which increases the energy by an amount
$\delta E$. Whatever the nature of the object that we add, this
cannot increase the entropy by more than it would have increased
had we added this energy as heat. Since we consider a small change
in the configuration, the first law yields
\begin{equation} \label{enbound}
\delta S \le \delta S_{max} = {\delta E \over T}.
\end{equation}
We see that the process of adding a small object saturates this
bound, at least to first order in small quantities. If on the
other hand we considered finite changes, we would still expect the
resulting $\delta S$ to satisfy the bound (\ref{enbound}), though
it would no longer be saturated.

At this point the reader may wish to see more carefully how the
above effect arises in an explicit example.  In particular, the
reader may question the step in which $\delta \rho$ is treated as
a small perturbation, which has so far been argued only on
physical grounds and without mathematical rigor\footnote{One can
do a bit better: An argument (due to Mark Srednicki) for this
result was given in the appendix of Ref. \cite{MMR} using a
standard trick of statistical mechanics.  However, this trick has
also not been rigorously justified.}.  
As a result, it was recently claimed\cite{Rumanov} that careful consideration of this issue
would find large corrections over the analysis of Ref. \cite{MMR}
and invalidate the result.   To study possible corrections in detail, we now turn in section \ref{calcs}
to a concrete example in the context of free bosons where all
quantities of interest may be computed explicitly.  We will find certain errors in Ref. \cite{Rumanov}, and show instead that the linear approximation is indeed highly accurate.

\subsection{An explicit example: Free Bosons}

\label{calcs}

Consider then a system of $n$ free boson fields.  Our ``objects'' will be merely the particles of such fields, with the understanding that the flavor of the particle is treated as unobservable.   That is, each 1-particle state will correspond to a microstate of our object, but any two such states which differ only by change in the flavor of particles will correspond to the same macrostate of our object.  Thus, we
wish to calculate the entropy of ``objects'' described, from the inertial
point of view, by a single-particle density matrix uniformly
distributed over the different fields. The inertial observer
therefore assigns the object an entropy $S_{inertial}= \ln n$.  This example
was briefly considered in Ref. \cite{MMR}, though only under the
assumption that $\delta \rho$ could be treated perturbatively.  An
attempt was then made in Ref. \cite{Rumanov} to analyze this key
assumption in more detail by computing the result exactly.  The
claim of that work was that discrepancies were found that were sufficient to invalidate
the conclusions of Ref. \cite{MMR}.  However, we will repeat this
analysis below and, after correcting some elementary mistakes in
the equations of Ref. \cite{Rumanov}, we will in fact find
support for the above observer-dependence of entropy and in particular for the accuracy of the linearized treatment used above.  

Our
notation and set up essentially follow that of Ref. \cite{MMR}.
To begin, we write the Minkowski vacuum for a system of $n$ free
bosonic fields in terms of the Rindler Fock space.  The result
is\cite{unruh}:
\begin{equation}
|0_M\rangle = \prod_{i=1}^n \prod_j
(1-e^{-\frac{\omega_j}{T}})^{\frac{1}{2}} e^{-\frac{\omega_j}{2T}
  a_{ijL}^{\dagger}a_{ijR}^{\dagger} } |0_{Rindler}\rangle .
\end{equation}
Here $i$ labels the different fields, $j$ labels a complete set of modes
$u_{jL}, u_{jR}$ of positive Rindler frequency $\omega_j$ for each field,
and the labels $R$ and $L$
refer respectively to the right and left Rindler wedges.  Each mode $u_{iL},u_{jR}$
of the $i$th field has an associated annihilation operator $a_{iL},a_{iR}$ which annihilates the Rindler
vacuum $|0_{Rindler}\rangle$ and satisfies a standard commutation relation of the form $a a^\dagger- a^\dagger a =1$.
The temperature $T$ associated with
the uniformly accelerating observer of interest is given by $T=\frac{a}{2\pi}
= \frac{\kappa}{2\pi}$ where
$a$ is the observer's proper
acceleration and $\kappa$ is the surface gravity of the boost Killing field $\xi$
normalized on our observer's worldline.

Similarly, the annihilation operator $a_{iM}$ for a Minkowski mode
$u_M$ of the $i$th field can on general principles\cite{unruh} be
expressed in the form:
\begin{eqnarray}
\label{Bog}
a_{iM} &=& \sum_{j}  \Bigl[ (u_M,u_{jR})( a_{ijR} -
e^{-\frac{\omega_j}{2T}} a_{ijL}^{\dagger})
\nonumber \\ &+& (u_M,u_{jL})( a_{ijL} -
e^{-\frac{\omega_j}{2T}} a_{ijR}^{\dagger})
\Bigr],
\end{eqnarray}
where $(u,v)$ is the Klein-Gordon inner product.  For simplicity
we might suppose that we choose a mode $u_M$ with no support on
the invisible Rindler wedge (say, the left one) and for which
$(u_M,u_{jR})$ is well modelled by a delta-function; we will
return to the general case later in the subsection.  In
particular, this simplification means that the Rindler frequency
$\omega$ of $u_M$ is reasonably well
defined\footnote{\label{foot}If the object is well localized and
located near our Rindler observer at some time, then $\omega$ is
also the frequency measured by the inertial observer.}.  For such
a case the above Bogoliubov transformation becomes
\begin{equation}
\label{trunc}
a_{iM} = \frac{1}{\sqrt{1-e^{-\frac{\omega}{T}}}}( a_{iR} -
e^{-\frac{\omega}{2T}} a_{iL}^{\dagger}),
\end{equation}
where the normalization of $a_{iM}$ is fixed by the commutation
relation. Here the Rindler operators refer to the one relevant
pair of Rindler modes (and $i$ continues to label flavors of fields). Note that since the we consider a free
field, all other modes are decoupled and can be dropped from the
problem.  Thus, we simplify the discussion below by considering
only the part of the Rindler Fock space associated with this pair
of modes.  Tracing over the invisible wedge yields 
\begin{equation}
\rho_{R0} = (1 - e^{-\omega/T})^n e^{-N\omega/T},
\end{equation}
where $N = \sum_i a^\dagger_{iR}  a_{iR} $ is the operator associate with the {\it  total} number of particles without regard to flavor.  

Since $a_{iM}$ acts in the Hilbert space describing only the $i$th field, it is
convenient to describe the action of $a_{iM}$ on the vacuum
$|0_{iM} \rangle$ for this particular field alone.  One sees that the
properly normalized Minkowski one-particle state is given
in terms of an infinite number of Rindler excitations
\begin{equation}
a_{iM}^{\dagger} |0_{iM}\rangle = (1-e^{-\frac{\omega}{T}}) \sum_k
e^{-\frac{k \omega}{2T}} \sqrt{k+1} |k, k+1\rangle_i,
\end{equation}
where $|k,k+1 \rangle_i$ denotes the state of field $i$ having $k$ Rindler
excitations in the right wedge and $k+1$ Rindler excitations in the left wedge.
Here we consider only the factor in the Hilbert space that describes
the modes appearing in (\ref{trunc}).

Thus, if we return to our Minkowski density matrix
\begin{eqnarray}
\label{dm}
\rho_M &=& {1 \over n} \sum_{i=1}^n |1_M;i \rangle \langle1_M;i| \nonumber \\ &=& {1
  \over n} \sum_{i=1}^n (a_{iM}^{\dagger} |0_M\rangle \langle0_M|
  a_{iM}),
\end{eqnarray}
tracing over the invisible Rindler wedge will give the desired
result. As in Ref. \cite{MMR}, a short calculation gives
\begin{equation}
\rho_{R1} =  {1 \over n} \sum_i \ (1-e^{-\omega/T})^n \sum_{k_1,...,k_n}
e^{-(\omega/T)
   \sum_j k_j} \ (e^{\omega/T}-1) k_i |k_1 \ldots k_n \rangle \langle k_1
\ldots k_n  |,
\end{equation}
where again only the part of the state associated with the Rindler
mode of interest has been displayed.  Here the notation $|k_1
\ldots k_n \rangle$ denotes the state with $k_j$ particles of type
$j$ in this mode of Rindler energy $\omega$ in the visible Rindler wedge.  One
can write this result in terms of the matrix elements between a
complete set of modes as
\begin{equation}
\rho_{R1} = \sum_{\vec{k}} (\rho_{R1})_{\vec{k}, \vec{k}} |k_1
\ldots k_n \rangle \langle k_1 \ldots k_n  |,
\end{equation}
where
\begin{equation}
(\rho_{R1})_{\vec{k}, \vec{k}} = \left[ (e^{\omega/T}-1) {1 \over n}
\sum_{i=1}^n
   k_i \right] (\rho_{R0})_{\vec{k}, \vec{k}},.
\end{equation}
and $\vec k = \{k_1, \ldots , k_n\}$.  As a result, one sees that
\begin{equation}
\label{simple}
\rho_{R1} = \frac{e^{\omega/T}-1}{n} N \rho_{R0}.
\end{equation}
This expression is the one given in Ref. \cite{MMR}.  However, in
Ref. \cite{Rumanov} $\sum_i k_i$ has been replaced by $k_i$ in
what appears to be a typographic mistake (as the left hand side is
clearly independent of the choice of any particular field $i$).
Note that $\rho_{R1}$ is diagonal in the standard basis. One can
check explicitly that $\rho_{R1}$ is properly normalized so that
its trace is 1.

The change in the density matrix is thus $\delta \rho = \rho_{R1}
- \rho_{R0}$.   For non-interacting particles, the Hamiltonian is
$H = \omega N$, so the average change in energy is
\begin{equation}
\label{dE}
E_{acc} := \delta E = \omega \mbox{Tr} (N\delta \rho) =
\frac {\omega}{1-  e^{-\frac{\omega}{T}}}.
\end{equation}
Note that (\ref{dE}) requires no approximation that $\delta \rho$
is in any sense ``small.''

As remarked in Ref. \cite{MMR}, this result is already of
interest. Note that for $\omega \gg T$ and under the conditions of
footnote (\ref{foot}), one finds $\delta E = E_{inertial} \approx
\omega$. On the other hand, for $\omega \ll T$ the background of
objects in the thermal bath leads to an amplification reminiscent
of the effect of stimulated emission.  In particular, for $\omega \ll T$ we have added (on average) far more than one object (in fact, $\frac{1}{1-e^{-\omega/T}}$ objects) 
to the thermal bath in constructing $\rho_{R1}$ from $\rho_{R0}$.

However, the quantity of real interest is $\delta S =Tr (\delta
\rho_{R0} \ln \rho_{R0}) - Tr (\delta \rho_{R1} \ln \rho_{R1})$.  
Let us therefore calculate the entropies $S_{R0}$ and $S_{R1}$ of $\rho_{R0}$ and $\rho_{R1}$.  Here we follow the calculation of Ref, \cite{Rumanov}, but correcting the mistake mentioned  above.   The entropy of $\rho_{R0}$ is
\begin{equation}
S_{R0}  =  - Tr \rho_{R0} \ln \rho_{R0} = n \left( \frac{\omega/T}{e^{\omega/T} -1} -  \ln(1 - e^{-\omega/T}) \right),
\end{equation}
since $\rho_{R0}$ is just a thermal ensemble of $n$ Harmonic oscillators.  On the other hand, we have
\begin{eqnarray}
\label{R1}
S_{R1} &=& -Tr\;\rho_{R1}\ln\rho_{R1} \cr
&=& Tr \left[ \rho_{R1}  \left(  \ln \frac{n}{(e^{\omega/T}-1)(1-e^{-\omega/T})^n} +
\frac{\omega}{T}N - \ln N  \right) \right] 
\cr
&=&
S_{R0} + \ln\frac{n}{e^{\omega/T}-1} + \frac{\omega/T}{1-e^{-\omega/T}} \cr
&-&
(1- e^{-\omega/T})^n {{e^{\omega/T} -1} \over n} \sum_{k_1,...,k_n=0}^\infty
(\sum_{i=1}^n k_i) e^{-(\sum_{j=1}^n k_j)\omega/T}
\ln (\sum_{m=1}^n k_m), \ \ \ \ \ \ \ \ \ \
\end{eqnarray}
where the final expression has been computed using (\ref{simple}).  The last term in the final expression is merely the last term in the second line written out in complete detail.  This allows the reader to compare with and correct Ref. \cite{Rumanov}.

This last term is difficult to evaluate explicitly, but one may estimate the result based on the fact that it is the expectation value of $-\ln N$ in the state $\rho_{R1}$.   Now, the expectation value of  $N$ in the thermal state $\rho_{R0}$ under these conditions is $\langle N \rangle_{R0} = \frac{n}{e^{\omega/T}-1} \gg 1$.   But $\rho_{R1}$ differs from $\rho_{R0}$ only by adding one object (using a Minkowski creation operator).  Since a Minkowski creation operator is a combination of a Rindler creation operator and a Rindler annihilation operator, the expectation value of $N$ in $\rho_{R1}$ will then lie in the range 
\begin{equation}
\frac{n}{e^{\omega/T}-1} -\delta < \langle N \rangle_{R1} < \frac{n}{e^{\omega/T}-1}+\delta,
\end{equation}
where $\delta$ is an unknown function of $\omega/T$ which measures the number of Rindler quanta created by the action of a Minkowski creation operator.  Based on eq. (\ref{dE}), it is natural to expect that $\delta$ contains a similar ``stimulated emission'' effect, so that
\begin{equation}
\label{enhance}
\delta = \frac{\alpha}{1-e^{-\omega/T}},
\end{equation}
with $\alpha$ independent of $\omega/T$.  We will find further evidence for this form from numerical calculations below.

Furthermore,  at large $n$ the distribution of $N$ will be sharply peaked about the mean $\langle N \rangle$, while we may approximate 
\begin{equation}
\label{est}
\ln(\langle N \rangle_{R0} + \delta) \approx \ln \langle N \rangle_{R0} + \frac{\delta}{\langle N \rangle_{R0}} = 
\ln \left( \frac{n}{e^{\omega/T}-1}\right)
+ \alpha \frac{e^{\omega/T}}{n}.
\end{equation}
Thus, one should be able to replace $-Tr [\rho_{R1} \ln N]$ with $\ln \left( \frac{n}{e^{\omega/T}-1}\right)$ up to a term of order $\frac{1}{ne^{-\omega/T}}$.  
Performing this substitution yields 
\begin{equation}
\label{result}
S_{R1} - S_{R0} = \frac{\omega/T}{e^{\omega/T} -1} +  O(\frac{1}{ne^{-\omega/T}}).
\end{equation}
Note that (\ref{result}) is indeed $E_{acc}/T = \frac{\langle H \rangle_{R0} -  \langle H \rangle_{R1}}{T}$ to leading order, as predicted by the linearized calculation.  In particular, for large $n$ the Rindler observer assigns a much smaller entropy than does the inertial observer.

\begin{figure}[ht]
\centerline{\epsfxsize=2.5 in\epsfbox{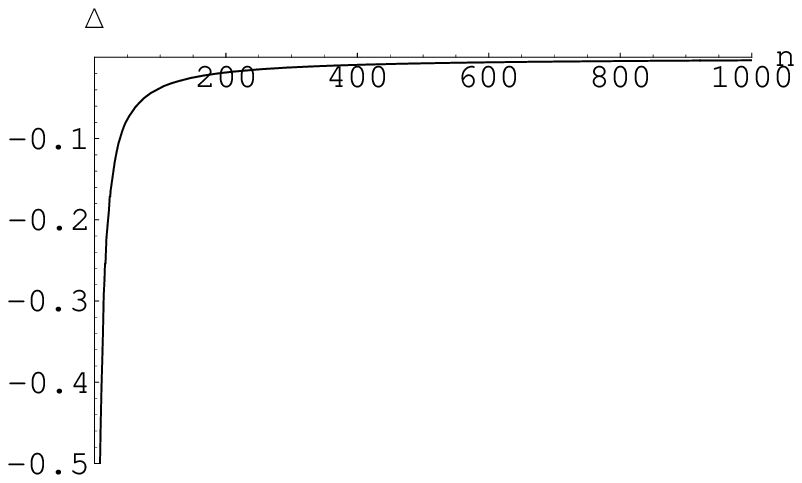} \ \ \epsfxsize=2.5in\epsfbox{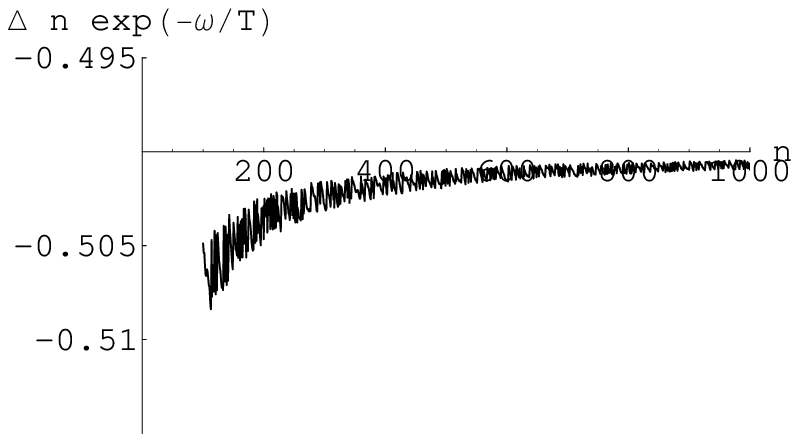}}
\caption{{\bf $\omega/T=2$.}  On the left we plot $\Delta$ vs. the number of fields $n$ up to $n=1000$, so that $ne^{-\omega/T}$ ranges up to roughly 135.  The result appears to converge to zero from below.  On the right, $\Delta$ has been scaled by $ne^{-\omega/T}$.  The product appears to converge to $-1/2$, further supporting the claim that $\Delta$ is $O([ne^{-\omega/T}]^{-1})$.  
\label{Fig1}}
\end{figure}

\begin{figure}[ht]
\centerline{\epsfxsize=2.5in\epsfbox{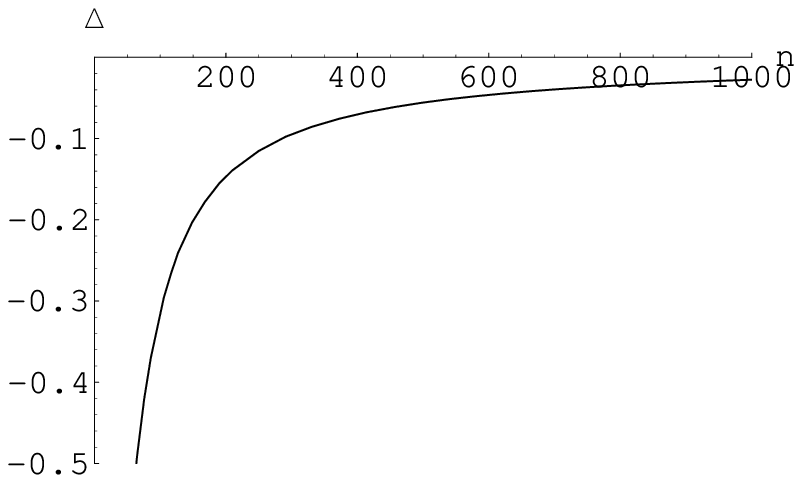} \ \ \epsfxsize=2.5in\epsfbox{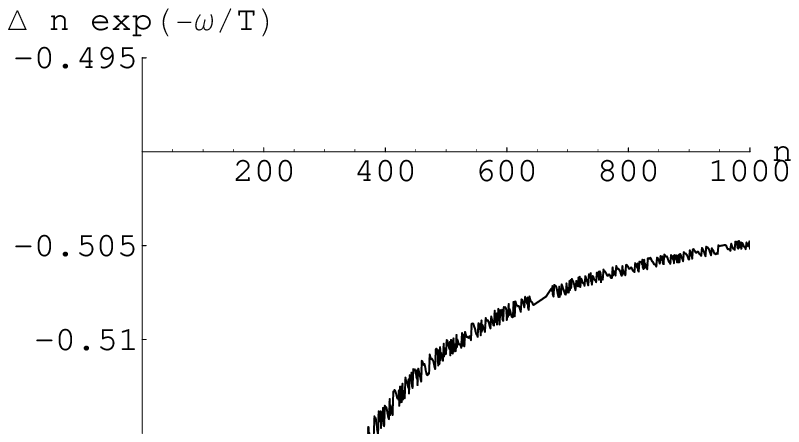}}
\caption{{\bf $\omega/T=4$.}  On the left we plot $\Delta$ vs. the number of fields $n$ up to $n=1000$, so that $ne^{-\omega/T}$ ranges up to roughly 18.  The result appears to converge to zero from below.  On the right, $\Delta$ has been scaled by $ne^{-\omega/T}$.  The product appears to converge to $-1/2$, further supporting the claim that $\Delta$ is $O([ne^{-\omega/T}]^{-1})$. \label{Fig4}}
\end{figure}

\begin{figure}[ht]
\centerline{\epsfxsize=2.5in\epsfbox{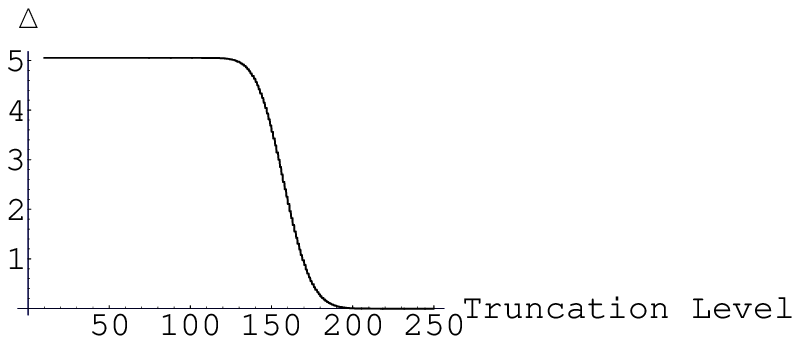} \ \ \ 
\epsfxsize=2.5in\epsfbox{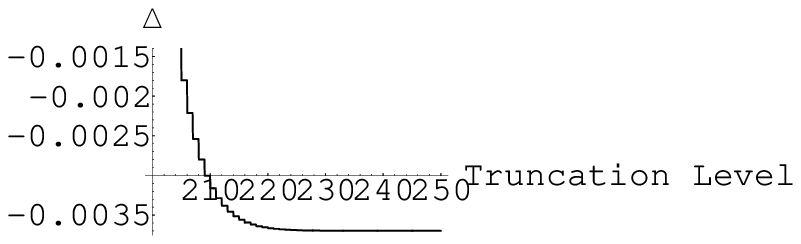}}
\caption{The accuracy of truncating the sum in (\ref{R1}) at a finite level is illustrated by plotting $\Delta$ against a range of truncation levels.  The figure on the right shows an expanded view of the region between truncation levels 200 and 250.  Note that a truncation level of 250 was used to generate Figs. (\ref{Fig1}) and (\ref{Fig4}).   \label{Fig2}}
\end{figure}

Lest the reader have some remaining doubt as to the accuracy of this approximation, some simple numerical calculations (performed using Mathematica) are included below.  Figs. (\ref{Fig1},\ref{Fig4})  each include two plots showing numerical computations of the error term $\Delta = S_{R1} - S_{R0} - E_{acc}/T$, which gives the discrepancy between the exact result and the linearized approximation.
These computations are done 
 by truncating the infinite sum in (\ref{R1}) at a fixed total number ($N$) of objects.  In Figs. (\ref{Fig1},\ref{Fig4}), this ``truncation level'' was taken to be $N=250$.  Fig. (\ref{Fig2}) shows the result of varying this truncation level and indicates that, for the parameters relevant to Figs. (\ref{Fig1},\ref{Fig4}),   the value of $\Delta$ has indeed stabilized by the time one reaches $N=250$.  

On the left side of Fig. (\ref{Fig1}), $\Delta$ has been plotted against the number of objects $n$ for the case $\omega/T=2$.   One can see that it rapidly approaches zero {\it from below}, indicating not only that $\delta S = E_{acc}/T$ in the limit $n \rightarrow \infty$, but also that $E_{acc}/T$ provides an upper bound on $\delta S$ as is required for more general preservation of the second law.
On the right in Fig. (\ref{Fig1}), an attempt has been made to gain better control over this residual error $\Delta$.  Since from (\ref{est}) one expects the error to scale with   $\frac{1}{ne^{-\omega/T}}$, we have plotted the ratio of $\Delta$ to this factor; i.e., the product $ \Delta ne^{-\omega/T}$.  This product does indeed appear to approach a constant in Fig. (\ref{Fig1}).  In particular, it appears to approach the value $-1/2$, corresponding to $\alpha = +1/2$ in (\ref{enhance}).    For comparison, Fig. (\ref{Fig4}) shows the corresponding plot for $\omega/T=4$ which shows similar results (and again suggests $\alpha = 1/2$).  

\section{Discussion}
\label{disc}

We have reviewed recent results pointing out loopholes in  arguments which had appeared to indicate that novel entropy bounds of the rough form of Eqs. (\ref{Bek}) or (\ref{holog}) were required for consistency of the generalized second law of thermodynamics.   Several mechanisms were identified that can protect the second law without (\ref{Bek}) or (\ref{holog}).  The primary such mechanisms are A) the high probability that a macrostate associated with a large entropy $S$ will be produced by thermal fluctuations, and thus in Hawking radiation and B) the realization that observers remaining outside a black hole associate a different (and, at least in interesting cases, smaller) flux of entropy across the horizon with a given physical process  than do observers who themselves  cross the horizon during the process.  In particular, this second mechanism was explored using both analytic and numerical techniques in a simple toy model.  We note that similar effects have been reported\cite{Alsing} for calculations involving quantum teleportation experiments in non-inertial frames.  Our observations are also in accord with general remarks\cite{Wald1,WaldR} that, in analogy with energy, entropy should be a subtle concept in General Relativity.

We have concentrated here on this new observer-dependence in the concept of entropy.  It is tempting to speculate that this observation will have further interesting implications for the thermodynamics of black holes.  For example, the point here that the two classes of observers assign different values to the entropy flux across the horizon seems to be in tune with the point of view (see, e.g., Refs. \cite{surface,s1,s2,s3,s4}) that the Bekenstein-Hawking entropy of a black hole does not count the number of black hole microstates, but rather refers to some property of these states relative to observers who remain outside the black hole.  Perhaps the observations above will play some role in fleshing out this point of view? 

In contrast, suppose that one remains a believer in, say, the Covariant Entropy Bound\cite{Bousso}, which purports to bound the flux of entropy across a null surface.  Since we have seen that this flux can depend on the choice of an observer, one may ask if this bound should apply to the entropy flux described by all observers, or only to one particular class of observers.  We look forward to discussions of these, and other such points, in the future.


\section*{Acknowledgments}

I am grateful to Djordje Minic, Simon Ross, and Rafael Sorkin for delightful collaborations leading to the results presented above and for other wonderful discussions.  I would also like to thank Bob Wald and Ted Jacobson for many fascinating lessons on black hole thermodynamics and black hole entropy.  Thanks also to Stefan Hollands for conversations relating to Refs. \cite{KW,FH}.  Much of the research on the observer dependence of entropy was conducted at the Aspen Center for Physics.  This work was supported in part by  NSF grant PHY-0354978 and by funds from the University of California.



\end{document}